\documentclass{webofc}
\usepackage[varg]{txfonts}   
\usepackage{upgreek}
\usepackage{lineno}
\modulolinenumbers[1]
\begin{document}


%
\title{Rescattering effects on resonances production in small systems with ALICE at the LHC}

\author{\firstname{Antonina} \lastname{Rosano}\inst{1,2}\fnsep\thanks{\email{arosano@unime.it}} on behalf of the ALICE Collaboration
}

\institute{Dipartimento MIFT, Università di Messina - Messina, Italy
\and   INFN, Sezione di Catania - Catania, Italy}

\abstract{Recent multiplicity-dependent analyses of pp and p--Pb collision data have revealed that particle production shares similar features with that in heavy-ion collisions. Studies using resonances could help to understand the possible onset of collective-like phenomena and the presence of a hadronic phase in small collision systems. Measurements of the differential yields of resonances with different lifetime, mass, quark content, and quantum numbers could enable understanding the mechanisms that influence the shape of particle momentum spectra, lifetime of the hadronic phase, strangeness production, parton energy loss, and collective effects. New ALICE results on various hadronic resonances in small collision systems at Large Hadron Collider (LHC) energies, including the multiplicity dependence measurements of $\Lambda$(1520) and K$^*$(892)$^{\pm}$ and the production of $\upphi$-meson pairs are presented here. Results will be also compared with model calculations.
}
\maketitle

\section{Introduction}
\label{intro}
Hadronic resonances are useful tools to characterize the late stage evolution (hadronic phase) of the system created by heavy-ion collisions at ultra-relativistic energies. In such collisions the critical values of temperature ($T_{\rm c} \simeq 160$~MeV) and energy density ($\epsilon_{\rm c} \simeq 1$~Gev/fm$^3$) can be reached with the consequent formation of a deconfined state of free quarks and gluons, known as quark--gluon plasma (QGP). Resonances with a lifetime comparable to that of the fireball ($\sim$ 1-10 fm/$c$ at LHC energies) may be sensitive to the competitive rescattering and regeneration mechanisms that could lead to a suppression or an enhancement of the resonance measured yield, respectively. The suitable conditions for QGP formation are expected to be reached only in heavy-ion collisions, with data from small systems (pp and p--Pb) collected only as a reference baseline. However recent studies of pp and p--Pb collisions at the LHC for events with high charged-particle multiplicities have shown patterns that are reminiscent of phenomena observed in A--A collisions. For example the strangeness enhancement~\cite{se}, the hardening of hadron $p_{\rm T}$ spectra~\cite{pp_hard1, pp_hard2, pPb_hard1, pp_K*0}, and the hint of suppression with increasing multiplicity of the ratios of short-lived resonances to long-lived hadrons~\cite{pp_K*0} have been observed even in small collision systems. The presence of typical features of heavy-ion collisions also in pp and p--Pb collisions is an intriguing observation whose interpretation could lead to new perspectives in the theoretical foundations behind QGP formation and to a better understanding of the mechanisms involved.

\section{Methodology}
\label{method}
ALICE (A Large Ion Collider Experiment~\cite{ali}) is one of the large experiments installed at the LHC at CERN. 
A complete description of the ALICE apparatus and its performance can be found in~\cite{ali, ali2}. The main sub-detectors involved in resonance analyses are the Inner Tracking System (ITS), the Time Projection Chamber (TPC), the Time-Of-Flight detector (TOF), and the V0A and V0C scintillators. The ITS and the TPC are used for primary vertex determination, tracking and particle identification (PID) via the measurement of the particle specific energy loss. The TOF detector is designed for PID measuring particle time of flight, and the V0A ($2.8 < \eta < 5.1$) and V0C ($-3.7 < \eta < -1.7$) hodoscopes are used for triggering and selecting events based on the charged particle multiplicity at forward rapidities. Resonances are reconstructed via invariant mass distribution of the hadronic decay products. The shape of the uncorrelated background is estimated by the event-mixing or like-sign technique. After combinatorial background subtraction, the invariant mass distribution is then fitted with a polynomial function to describe the residual background and with a Breit-Wigner, a Voigtian or a Gaussian function to describe the signal. Corrective factors as geometrical acceptance and detector efficiency, branching ratio, trigger selection efficiency, and signal-loss factor are applied to the raw yields in order to estimate the final yields.

\section{Results and discussion}
\label{results}
\begin{figure}[h!]
    \centering
    \includegraphics[scale=0.18]{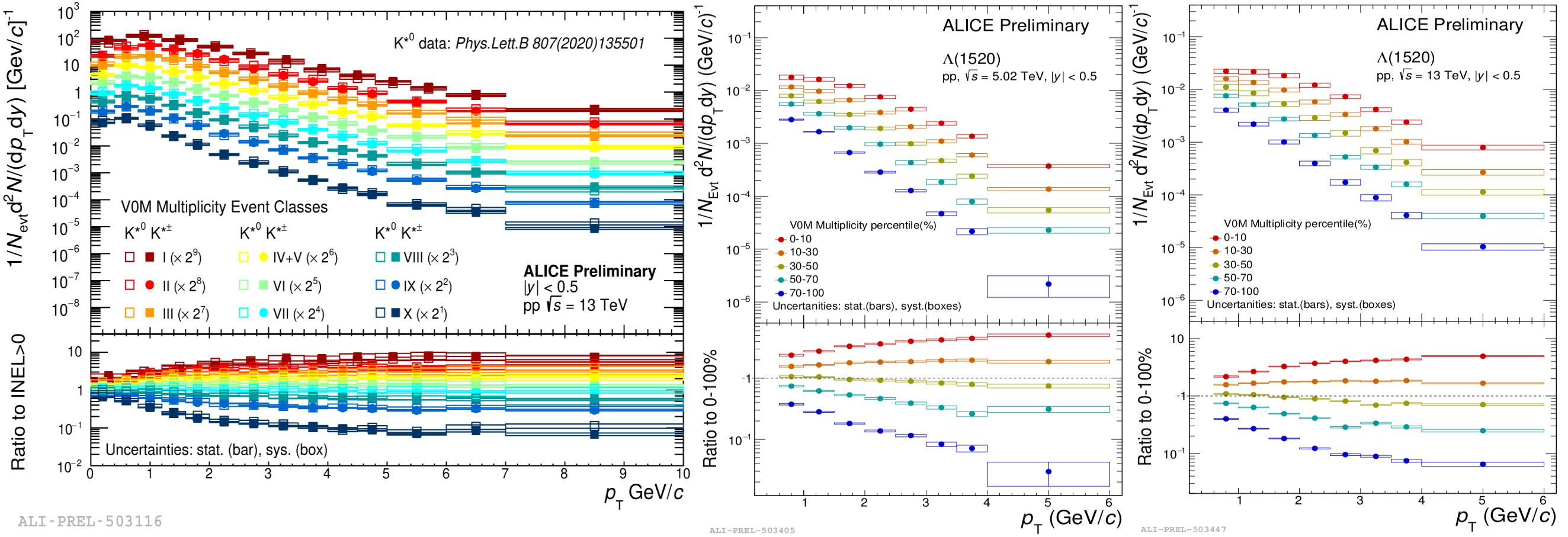}
     \put(-285,111){\bfseries (a)}
    \put(-160,111){\bfseries (b)}
    \put(-60,111){\bfseries (c)}
    \caption{(Colour online) The $p_{\rm T}$-differential distributions of K$^{*\pm}$ in pp collisions at $\sqrt{s} = 13$~TeV {\bfseries (a)}, $\Lambda(1520)$ at $\sqrt{s} = 5.02$  {\bfseries (b)} and 13~TeV {\bfseries (c)}. Lower panels show the ratios of the $p_{\rm T}$ distributions to the inclusive (INEL$>$0) spectra. The K$^{*\pm}$ data are also compared to K$^{*0}$ measurements~\cite{pp_K*0}.}
    \label{fig:spectra}
\end{figure}

ALICE has collected data from several collision systems at different LHC energies, measuring the production of a large set of hadronic resonances~\cite{pp_K*0, rho, PbPbKandPhi, pPbKandPhi, pp7SigandXi, pPbSigandXi, PbPbSigma, Lambda_pp_pPb, Lambda_PbPb}. The latest results from the multiplicity ($\langle$d$N_{\rm ch}$/d$\eta\rangle_{|\eta|< 0.5}$) dependent analyses performed at $|y|< 0.5$ of K$^*(892)^{\pm}$ in pp collisions at $\sqrt{s} = 13$~TeV and $\Lambda(1520)$ in pp collisions at $\sqrt{s} = 5.02$ and 13~TeV, and of $\upphi$-meson pair in minimum bias pp collisions at $\sqrt{s} = 7$~TeV are discussed here.\\
Both K$^{*\pm}$ and $\Lambda(1520)$ $p_{\rm T}$-differential distributions get harder for $p_{\rm T} \lesssim 5$~GeV/$c$ going from low to high multiplicity classes (Fig.~\ref{fig:spectra}). The trend is qualitatively similar to the behaviour observed in Pb--Pb collisions that it is usually attributed to a collective radial expansion of the system (radial flow), although the colour reconnection mechanism seems to mimic collective-like effects in small systems~\cite{CR}.

\begin{figure}[h!]
    \centering
    \includegraphics[scale=0.2]{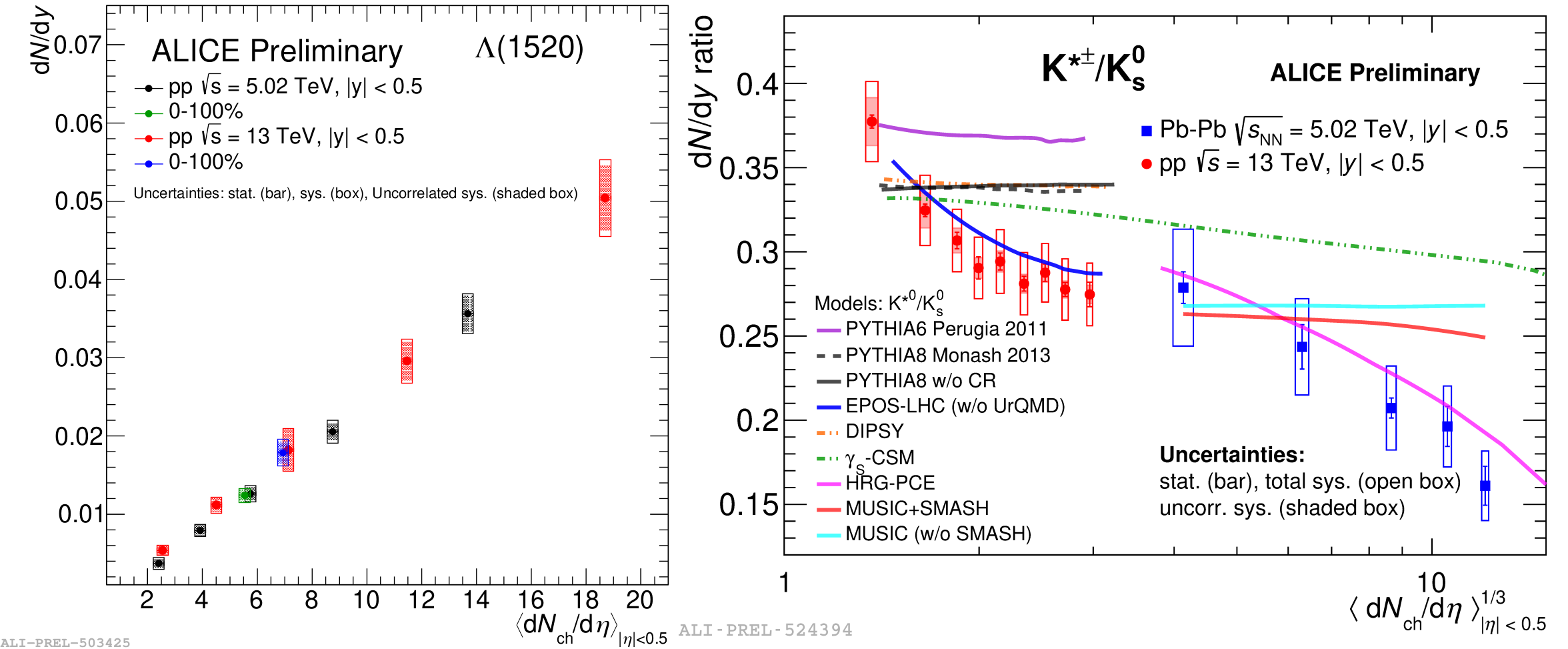}
     \put(-225,25){\bfseries (a)}
    \put(-140,128){\bfseries (b)}
    \caption{(Colour online) $p_{\rm T}$-integrated yields of $\Lambda(1520)$ as a function of $\langle$d$N_{\rm ch}$/d$\eta\rangle_{|\eta|< 0.5}$ in pp collisions at $\sqrt{s} = 5.02$  and 13~TeV {\bfseries (a)}. Ratios of $p_{\rm T}$-integrated yields of K$^{*\pm}$/K$^0_{\rm S}$ in pp at $\sqrt{s} = 13$~TeV and in Pb--Pb collisions at $\sqrt{s_{\rm NN}} = 5.02$~TeV as a function of multiplicity. Measurements are also compared with predictions from common event generators~\cite{epos, hrg, pythia6, pythia8, pythia8CR, dipsy, gamma_csm, music} {\bfseries (b)}. }
    \label{fig:dndy}
\end{figure}

In addition, these results have many qualitative similarities to those reported for longer lived hadrons, where a hardening at low $p_{\rm T}$ of the $p_{\rm T}$-differential distributions have been observed too, and are consistent with previous measurements of K$^{*0}$ and $\Lambda(1520)$ both in pp and p--Pb collisions~\cite{pp_hard1, pp_hard2, pPb_hard1, pp_K*0, pPbKandPhi, Lambda_pp_pPb}.\\
\phantom{x}\hspace{2ex}Figure~\ref{fig:dndy}a shows the $\Lambda(1520)$ $p_{\rm T}$-integrated yields as a function of the charged particle multiplicity density in pp collisions. The yields exhibit a linear increase with increasing multiplicity and results at $\sqrt{s} = 5.02$ and 13~TeV are consistent. Therefore, as observed also for other hadron species~\cite{pp_hard1, pp_K*0}, the particle production rate is independent of collision energy and scales with the event multiplicity.
The K$^{*\pm}$/K$^0_{\rm S}$ ratios are shown in Fig.~\ref{fig:dndy}b. The ratio of resonance $p_{\rm T}$-integrated yields to the production of their non-resonant hadronic states, is an important tool to verify the presence of a suppression in resonances production and its dependence on the system size. The K$^{*\pm}$/K$^0_{\rm S}$ ratio is suppressed in central Pb--Pb collisions with respect to peripheral Pb--Pb and pp collisions. This is an expected behaviour in Pb--Pb collisions since the measured yield of short-lived resonances as  K$^{*\pm}$ ($\tau~\simeq~4$~fm/$c$) can be affected by rescattering effects during the hadronic phase. However an intriguing suppression of the K$^{*\pm}$/K$^0_{\rm S}$ ratio is observed from low to high multiplicity pp collision too, which is significant at the $7\sigma$ level taking into account the fraction of systematic uncertainties uncorrelated with multiplicity. The latest measurement confirms and improves the precision of previous K$^{*0}$ results~\cite{pp_K*0} due to lower systematic uncertainties. The measured values are compared with several model calculations. Among them, EPOS-LHC~\cite{epos} without the UrQMD hadronic afterburner and the hadron resonance gas model in partial chemical equilibrium (HRG-PCE)~\cite{hrg} provide the best description for pp and Pb--Pb data respectively, reproducing the decreasing trend.

\begin{figure}[h!]
    \centering
    \sidecaption
    \includegraphics[scale=0.22]{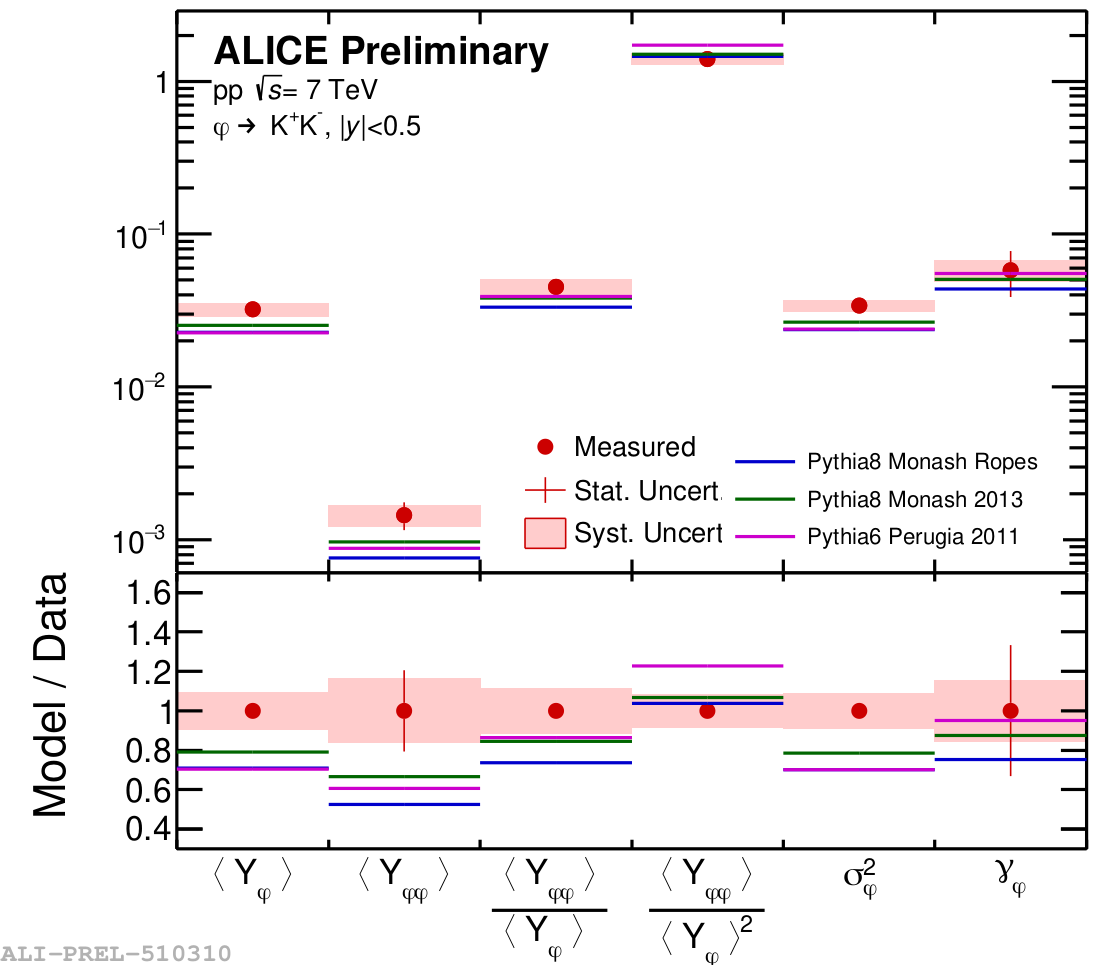}
    \caption{(Colour online). Inclusive $\upphi$ ($\langle {\rm Y}_{\upphi} \rangle$) and double $\upphi$-meson yields ($\langle {\rm Y}_{\upphi\upphi} \rangle$) with their combinations. The key observable $\gamma_{\upphi}$ is also shown. Results are compared with the most commonly used PYTHIA tuned models (solid colored lines)~\cite{pythia6, pythia8, pythia8CR}. Lower panel shows the ratios of the model predictions to the measured data.}
    \label{fig:phi}     
\end{figure}

Figure~\ref{fig:phi} shows the results of the study on the double $\upphi$ production in pp collisions at $\sqrt{s} = 7$~TeV, performed in order to investigate strangeness production in small systems. In terms of statistical properties of the $\upphi$-meson production, the average yield ($\mu = \langle {\rm Y}_{\upphi} \rangle$) and the variance ($\sigma^2 = \langle {\rm Y}_{\upphi}^2 \rangle - \langle {\rm Y}_{\upphi} \rangle^2$) have been considered. Looking at the variance definition, the second term can be directly measured while the first one can be obtained via the $\upphi$ meson pair production ($\langle {\rm Y}_{\upphi\upphi} \rangle$):

\begin{equation}
    \langle {\rm Y}_{\upphi}^2 \rangle = 2\langle {\rm Y}_{\upphi\upphi} \rangle + \langle {\rm Y}_{\upphi} \rangle \Rightarrow \sigma^2 = 2\langle {\rm Y}_{\upphi\upphi} \rangle + \langle {\rm Y}_{\upphi} \rangle - \langle {\rm Y}_{\upphi} \rangle^2
\end{equation}

This information can be used to define $\gamma_{\upphi}$, a new parameter that describes the accordance of the production statistic with a Poissonian beahaviour:

\begin{equation}
    \gamma_{\upphi} = \frac{\sigma^2}{\mu} -1 = \frac{2\langle {\rm Y}_{\upphi\upphi}\rangle}{\langle {\rm Y}_{\upphi}\rangle} - \langle {\rm Y}_{\upphi}\rangle
\end{equation}

In particular, $\gamma_{\upphi} = 0$ is the expected value if the production follows a Poissonian distribution. The obtained results show $\gamma_{\upphi} > 0$ indicating that the production distribution is enhanced and not purely Poissonian.

\end{document}